\documentclass[prd,reprint,nofootinbib,superscriptaddress,preprintnumbers,showpacs]{revtex4-1}
\usepackage{amsfonts}
\usepackage{amssymb}
\usepackage{amsmath}
\usepackage{graphicx,color}
\usepackage{dcolumn}
\usepackage{epsfig}
\usepackage{bm}
\usepackage{bbm}
\usepackage{ulem}
\usepackage{hyperref}
\usepackage{lineno}

\usepackage{color}

\def\gtap{\ \raise.3ex\hbox{$>$\kern-.75em\lower1ex\hbox{$\sim$}}\ }
\def\ltap{\ \raise.3ex\hbox{$<$\kern-.75em\lower1ex\hbox{$\sim$}}\ }
\begin{document}
%\linenumbers

\title{
$X$ structures in $B^+\to J/\psi\,\phi\, K^+$
as one-loop and double-triangle threshold cusps
}
%%%%%%%%%%%%%%%%%%%% Authors %%%%%%%%%%%%%%%%%%%%%%%%%%%%%%%%%
\author{Satoshi X. Nakamura}
\email{satoshi@ustc.edu.cn}
\affiliation{
University of Science and Technology of China, Hefei 230026, 
People's Republic of China
}
\affiliation{
State Key Laboratory of Particle Detection and Electronics (IHEP-USTC), Hefei 230036, People's Republic of China}

\begin{abstract}
The LHCb data on $B^+\to J/\psi\phi K^+$
show four peaks and three dips in 
the $J/\psi\phi$ invariant mass distribution, and
the peaks are interpreted as
$X(4140)$, $X(4274)$, $X(4500)$ and $X(4685)/X(4700)$ resonance contributions.
Interestingly, 
{\it all} the peaks and dips are located at (or close to)
$D^*_{s}\bar{D}^{(*)}_{s}$, 
$D_{s0}^*(2317)\bar{D}^{(*)}_{s}$,
$D_{s1}(2536)\bar{D}^{(*)}_{s}$, and
$\psi'\phi$ thresholds.
These coincidences suggest a close connection between 
the structures and the thresholds, 
which however has not been seriously considered 
in previous theoretical studies on the $X$ structures.
In fact, if we utilize this connection and interpret 
the $X$ structures as common $s$-wave threshold cusps,
we face a difficulty:
$X(4274)$ and $X(4500)$ have
spin-parity that conflict with the experimentally determined ones. 
In this work, we introduce double triangle mechanisms that cause 
threshold cusps significantly sharper than 
the ordinary one-loop ones of the same spin-parity.
We demonstrate that all the $X$ and dip structures
are well
 described by a combination of one-loop and double-triangle threshold
 cusps, thereby proposing a novel interpretation of the $X$ and dip structures.
\end{abstract}

\maketitle

\section{introduction}

The $J/\psi\phi$ invariant mass ($M_{J/\psi\phi}$)
distribution of $B^+\to J/\psi\phi K^+$~\footnote{The charge conjugate
decays are implied throughout.}\footnote{
We follow the hadron naming scheme of Ref.~\cite{pdg}. 
For simplicity, however, 
$J/\psi$ and $\psi(2S)$ 
are often denoted 
by $\psi$  and $\psi'$, respectively.
We generically denote
$D_{s0}^*(2317)$ and $D_{s1}(2536)$
by $D_{sJ}^{(*)}$.
Charge indices are often suppressed.
}
shows structures, hinting the existence of exotic hadrons ($X$)
that are beyond the conventional 
$q\bar{q}$ and $qqq$
constituent quark structure.
After earlier analyses based on fitting only the $M_{J/\psi\phi}$
distribution~\cite{cdf,belle,cdf2,lhcb_old,cms,d0,babar,d02}, 
the LHCb Collaboration conducted a first six-dimensional amplitude analysis
and claimed four $X$ states along with their spin-parity ($J^P$)~\cite{lhcb_phi1,lhcb_phi2}:
$X(4140)$ and $X(4274)$ with $J^P=1^+$; 
$X(4500)$ and $X(4700)$ with $J^P=0^+$.
Recent higher statistics data confirmed 
these $X$ states, and added 
$1^+ X(4685)$,
$2^- X(4150)$, and 
$1^- X(4630)$~\cite{lhcb_phi}.
Moreover, structures in 
$M_{J/\psi K^+}$ distribution were interpreted with
$1^+ cu\bar{c}\bar{s}$ tetraquarks
$Z_{cs}(4000)^+$ and $Z_{cs}(4220)^+$;
see Table~\ref{tab:X}.
The $X$ states have been commonly interpreted 
as charmonium ($\chi_{cJ}$)~\cite{ortega1,dychen,wjdeng,oncala,molina,lcgui,Badalian,Ferretti,Ferretti2,mxduan,qflu}, 
hybrid~\cite{oncala}, and tetraquark ($cs\bar{c}\bar{s}$)~\cite{Stancu,maiani,rzhu,qflu,jwu,agaev,hxchen,zgwang1,zgwang2,zgwang3,cdeng,anwar,turkan,jwu2,zgwang4,yyang,zgwang5,cdeng2,Ghalenovi,ppshi,zgwang,xliu}.
Hadron molecule models were developed for
$X(4140)$~\cite{turkan,zmding,simonov} 
and 
$X(4274)$~\cite{jhe}.
See reviews~\cite{review_olsen,review_esposito,review_Karliner,review_Albuquerque,review_agaev}.
\begin{table}[b]
\renewcommand{\arraystretch}{1.4}
\tabcolsep=4.mm
\caption{\label{tab:X} $X$ and $Z_{cs}^+$
 from the LHCb analysis on 
$B^+\to J/\psi\phi K^+$~\cite{lhcb_phi};
$X\to J/\psi\phi$ and $Z_{cs}^+\to J/\psi K^+$.
}
\begin{tabular}{cc}\hline
   $J^P=1^+$ & $J^P=0^+$ \\ \hline
  $X(4140)$\ $X(4274)$\ $X(4685)$ &   $X(4500)$\ $X(4700)$\\
 $Z_{cs}(4000)^+$\ $Z_{cs}(4220)^+$ &  \\\hline
\end{tabular}

\end{table}

It is recognized~\cite{xhliu,xkdong2}
that the $X(4274)$ and $X(4500)$ 
peak positions are virtually at the 
$D_{s0}^*(2317)\bar{D}_s$ and 
$D_{s1}(2536)\bar{D}_s$ 
thresholds, respectively, 
and $X(4700)$ and $X(4685)$ are at the
$\psi'\phi$ threshold; see Fig.~\ref{fig:comp-data}.
The $X(4140)$ structure 
is close to the 
$D_{s}^*\bar{D}_s$ threshold.
Furthermore, 
three dip structures have their lowest points at the 
$D^{*}_{s}\bar{D}^*_s$ and 
$D^{(*)}_{sJ}\bar{D}^*_s$
thresholds.
This seems to suggest that the X and dip structures
are associated with openings of the
$D^{*}_{s}\bar{D}^{(*)}_s$ and 
$D^{(*)}_{sJ}\bar{D}^{(*)}_s$
channels through kinematical effects such as
threshold cusps and triangle singularities~\cite{ts_review}.

Indeed, the LHCb confirmed that the 
$X(4140)$ structure can be described with a 
$D^*_{s}\bar{D}_s$ threshold cusp,
albeit using a rather small cutoff in form factors~\cite{lhcb_phi2,swanson}~\footnote{
This LHCb's finding should be viewed with a caution 
since a small cutoff makes a cusp significantly sharper
by suppressing the high momentum contribution.
}.
Similarly, Liu studied 
triangle diagrams that cause
$D^*_{s}\bar{D}_s$ and 
$\psi'\phi$ threshold cusps, and 
found $X(4140)$- and $X(4700)$-like enhancements, respectively~\cite{xhliu}.
Dong et al. also suggested that 
$X(4140)$ could be caused by 
a $D^*_{s}\bar{D}_s$ virtual state and the associated 
threshold cusp~\cite{xkdong2}.
$X(4140)$ as the kinematical effect may be supported by 
a lattice QCD that found 
no 
$J^{PC}=1^{++}$ 
$cs\bar{c}\bar{s}$ state
below 4.2~GeV~\cite{Padmanath}.
On the other hand, $X(4274)$ [$X(4500)$] as an
$s$-wave $D_{s0}^*(2317)\bar{D}_s$ [$D_{s1}(2536)\bar{D}_s$]
threshold cusp
has $J^P$ that conflicts with the experimentally determined ones~\cite{lhcb_phi2,lhcb_phi}~\footnote{
\label{foote4}
Our present analysis assumes that $J^P$ of the $X$ structures 
determined by the LHCb~\cite{lhcb_phi2,lhcb_phi} are correct. 
It is noted, however, that 
the LHCb's $J^P$ determination is based on fitting the $X$ structures with
Breit-Wigner models and thus is not model-independent. 
If the $X$ and dip structures are described with more complicated
mechanisms that might involve kinematical effects, 
it is unclear whether $J^P$ of the $X$ structures remain
unchanged. }.
Non $s$-wave threshold cusps
from one-loop diagrams are unlikely either, 
since they should be suppressed~\cite{xhliu}.
Thus, until the present work, there exists no explanation of 
$X(4274)$ and $X(4500)$ based on kinematical effects.

\begin{figure*}
\begin{center}
\includegraphics[width=1\textwidth]{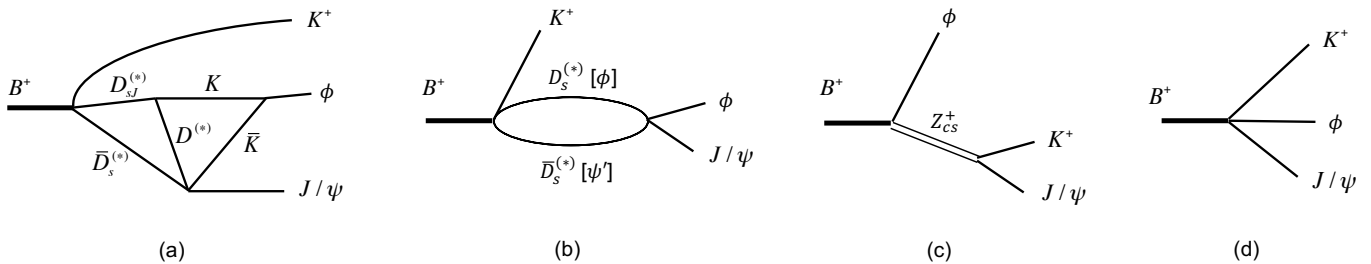}
\end{center}
 \caption{
$B^+\to J/\psi \phi K^+$ 
mechanisms:
(a) double triangle;
(b) one-loop;
(c) $Z_{cs}$ excitation;
(d) direct decay.
 }
\label{fig:diag}
\end{figure*}

Now let us assume negligibly small 
$D_{s0}^*(2317)\bar{D}_s^{(*)}$, $D_{s1}(2536)\bar{D}_s^{(*)}$
$\to J/\psi\phi$ 
transition strengths
caused by short-range (e.g., quark-exchange) interactions.
This assumption may seem reasonable for the $D_{s0}^*(2317)$ cases,
because previous theoretical studies~\cite{liu_Ds0,torres_Ds0,Cheung_Ds0,Yang_Ds0}
indicated a dominant $DK$-molecule component 
in $D_{s0}^*(2317)$.
Under this assumption, 
double triangle (DT) mechanisms of Fig.~\ref{fig:diag}(a)
should be the most important among those including
$D_{sJ}^{(*)}\bar{D}_s^{(*)}$.
The DT mechanisms are worthwhile studying
to understand the $X$ and dip structures and their locations. 
The DT mechanisms 
cause threshold cusps that are significantly sharper than
ordinary one-loop ones with the same $J^P$.
This is because the DT is close to causing the leading
kinematical singularity. 
Thus, the DT can generate $X$-like and dip structures
at the $D_{sJ}^{(*)}\bar{D}_s^{(*)}$ thresholds.

In this paper,
we develop a $B^+\to J/\psi \phi K^+$
decay model.
The model includes 
the DT mechanisms that cause 
enhanced threshold cusps at 
the $D_{sJ}^{(*)}\bar{D}_s^{(*)}$ 
thresholds.
$D_{s}^{*}\bar{D}_s^{(*)}$ and $\psi'\phi$
threshold cusps are also generated by
one-loop mechanisms.
We first examine singular behaviors of the DT amplitudes.
We then analyze the $M_{J/\psi\phi}$ distribution from the LHCb.
Since one-dimensional analysis 
would not reliably 
extract partial wave amplitudes
or determine spin-parity of resonances,
this is not our intention.
The present one-dimensional analysis
keeps $J^P$ of $X$ from the LHCb analysis.
Under this constraint, 
we demonstrate that all the $X$ and dip structures 
can be well described with the threshold cusps.
The purpose of this work is to 
propose a novel interpretation of the $X$ and dip structures
in this way.
\section{model}
In describing $B^+\to J/\psi \phi K^+$, 
we explicitly consider mechanisms 
that generate the structures in the $M_{J/\psi\phi}$ distribution
through kinematical effects or resonance excitations; others are
subsumed in contact mechanisms.
Thus we consider diagrams shown in Fig.~\ref{fig:diag}.
To derive the corresponding amplitudes,
we write down effective Lagrangians of relevant hadrons and their 
matrix elements,
and combine them following the time-ordered perturbation theory.
We consider the DT diagrams [Fig.~\ref{fig:diag}(a)] that include
$p$-wave pairs of 
\begin{eqnarray}
D_{s0}^*(2317)\bar{D}_s(1^+), \ \
D_{s0}^*(2317)\bar{D}^*_s(0^+,1^+), \nonumber\\
D_{s1}(2536)\bar{D}_s(0^+,1^+), \ \
D_{s1}(2536)\bar{D}^*_s(0^+,1^+),
\label{eq:ddpair}
\end{eqnarray}
where $J^P$ of a pair is indicated
in the parenthesis.
In principle, 
more quantum numbers are possible such as 
$J^P$ from $s$-wave pairs and $J^P=2^+$ from $p$-wave pairs
which the LHCb did not find relevant to the $X$ structures. 
While the kinematical effects can generate structures in lineshapes
almost model-independently, it is the dynamics 
that determines the
strength of the kinematical effects. 
Since the relevant dynamical information is scarce, we need to rely on the LHCb
analysis to select the quantum numbers to take into account in the
model. 
Most phenomenological models share 
this limitation of predicting quantum numbers relevant to the process. 
We assume that contributions from the other quantum numbers
are relatively minor and can be absorbed by mechanisms included in the
model. 
We also do not consider $D_{s1}(2460)\bar{D}_s^{(*)}$ pairs
since their threshold cusps are either not clear in the data or 
replaceable by a $D_{s0}^*(2317)\bar{D}^*_s$ threshold cusp.
The one-loop diagram [Fig.~\ref{fig:diag}(b)] includes 
$s$-wave pairs of 
$D_s^*\bar{D}_s(1^+)$,
$D_s^*\bar{D}^*_s(0^+)$, and 
$\psi'\phi(0^+,1^+)$;
$D_s^*\bar{D}^*_s(1^+)$ is not included since 
$D_s^*\bar{D}^*_s(1^+)\to J/\psi\phi(1^+)$ is forbidden by the
$C$-parity conservation.
We denote the DT and one-loop amplitudes by 
$A^{\rm DT}_{D_{sJ}^{(*)}\bar{D}_s^{(*)}(J^P)}$ and 
$A^{\rm 1L}_{D_{s}^{(*)}\bar{D}_s^{(*)}(J^P)}$ 
[or $A^{\rm 1L}_{\psi'\phi(J^P)}$], respectively.

We consider $Z_{cs}$
excitations [Fig.~\ref{fig:diag}(c)]
since the 
data~\cite{lhcb_phi} shows their effects on the $M_{J/\psi\phi}$
distribution. In particular, $Z_{cs}(4000)$ seems to enhance
the $X(4274)$ peak through an interference.
The LHCb presented the $Z_{cs}(4000)$ and $Z_{cs}(4220)$ properties. 
Meanwhile, coupled-channel
analyses~\cite{zcs_model1,zcs_model2,zcs_model3} 
found virtual states below the $D_s^{(*)+}\bar{D}^{*0}$
thresholds that may be identified with 
$Z_{cs}(4000)$ and $Z_{cs}(4220)$.
The $D_s^{(*)+}\bar{D}^{*0}$ threshold cusps 
enhanced by the virtual states
can fit the $M_{J/\psi K^+}$ distribution of $B^+\to J/\psi\phi K^+$~\cite{zcs_model1}.
Thus we examined the above two options.
We use a Breit-Wigner form without addressing the $Z_{cs}$ internal structures.
To simulate the $D_s^{(*)+}\bar{D}^{*0}$ threshold cusps,
two $Z_{cs}$ masses are 3975~MeV and 4119~MeV from 
the $D_s^{(*)+}\bar{D}^{*0}$ thresholds;
$Z_{cs}$ widths are set to be 100~MeV (constant width values); 
see Eqs.~(\ref{eq:Zcs1}) and (\ref{eq:Zcs2}) for formulas.
For each $Z_{cs}$,
we use a $p$-wave
$B^+\to Z_{cs}\phi$ decay vertex which contributes to the 
$1^+$ $J/\psi\phi$ final state.
Our fits visibly favored
the threshold-cusp-based $Z_{cs}$;
we thus use them hereafter.

All the other mechanisms 
such as non-resonant and $K^{(*)}_J$-excitations
are simulated by two independent direct decay mechanisms [Fig.~\ref{fig:diag}(d)] 
creating $J/\psi\phi(0^+,1^+)$.
We consider $J/\psi\phi(0^+,1^+)$ partial waves.
Although 
the LHCb amplitude analysis found
resonances in $1^-$ and $2^-$ partial waves, 
their contributions are rather small 
in the $M_{J/\psi\phi}$ spectrum.
We confirmed that the $1^-$ and $2^-$
resonance contributions 
only marginally improved our fits;
we thus do not consider them.

The DT and one-loop diagrams are respectively initiated by
$B^+\to D_{sJ}^{(*)}\bar{D}^{(*)}_sK^+$ and 
$B^+\to D_{s}^{*}\bar{D}^{(*)}_sK^+$ that
may be dominated by color-favored quark mechanisms.
Although charge analogous 
$B^+\to \bar{D}_{sJ}^{(*)}{D}^{(*)}_sK^+$ and
$B^+\to \bar{D}_{s}^{*}{D}^{(*)}_sK^+$
generally have independent decay strengths,
the corresponding DT and one-loop amplitudes have the same
singular behaviors as the original ones.
Thus we do not explicitly consider 
the charge analogous processes, but 
their effects and projections onto positive $C$-parity states
are understood to be taken into account in
coupling strengths of the considered processes.

We present amplitude formulas for representative cases;
see Appendix~\ref{app1} for
complete formulas.
We use the particle mass and width values from Ref.~\cite{pdg}
unless otherwise specified,
and denote the energy, momentum, and polarization vector of a particle $x$ by
$E_x$, $\bm{p}_x$, and $\bm{\epsilon}_x$, respectively.
A DT diagram
[Fig.~\ref{fig:diag}(a)]
that includes $D_{s0}^*(2317)\bar{D}_s(1^+)$ 
consists of 
four vertices such as 
$B^+\to D_{s0}^*\bar{D}_sK^+$,
$D_{s0}^*\to DK$,
$D\bar{D}_s\to J/\psi\bar{K}$, and
$K\bar{K}\to\phi$
given as
\begin{eqnarray}
&& c_{D_{s0}^*\bar{D}_s(1^+)}\,
\bm{p}_{\bar{D}_s}\cdot\bm{p}_K\,
 F_{D_{s0}^*\bar{D}_s K,B}^{11}\ , \\
 &&c_{DK,D_{s0}^*}\, f_{DK,D_{s0}^*}^{0} \ , \\
&& c^{1^-}_{\psi\bar{K},D\bar{D}_s}\,
i(\bm{p}_{\bar{K}\psi}\times \bm{\epsilon}_{\psi})\cdot 
\bm{p}_{D\bar{D}_s}\,
 f_{\psi\bar{K}}^{1} 
 f_{D\bar{D}_s}^{1}, \\
&& c_{K\bar{K},\phi}\,
\bm{p}_{\bar{K}K}
\cdot \bm{\epsilon}_\phi \,
 f_{K\bar{K},\phi}^{1} \ ,
\end{eqnarray}
respectively; $\bm{p}_{ab}\equiv \bm{p}_{a}-\bm{p}_{b}$.
We have introduced 
dipole form factors 
$F_{ijk,l}^{LL'}$,
$f_{ij}^{L}$, and $f_{ij,k}^{L}$ including a cutoff $\Lambda$.
We use a common cutoff value $\Lambda=1$~GeV in all form factors
unless otherwise stated.
We used a $p$-wave $D\bar{D}_s\to J/\psi\bar{K}$ interaction;
$s$-wave is forbidden by the spin-parity conservation~\footnote{
$s$-wave $D\bar{D}_s^*, D^*\bar{D}_s^{(*)}\to J/\psi\bar{K}$ 
interactions are allowed
in DT mechanisms including other $D_{sJ}^{(*)}\bar{D}_s^{(*)}$ pairs.
However, such DT amplitudes are suppressed due to their tensor
structures.
See a discussion above Eq.~(\ref{eq:a27}) in Appendix~\ref{app1}.}.
The coupling $c_{K\bar{K},\phi}$ can be 
determined by the $\phi\to K\bar{K}$ decay width.
Experimental information for
the other couplings
($c_{D_{s0}^*\bar{D}_s(1^+)}$,
$c_{DK,D_{s0}^*}$,
$c^{1^-}_{\psi\bar{K},D\bar{D}_s}$)
are unavailable. 
Thus 
we determine their product, which is generally a complex value, 
by fitting the data. 
The DT amplitude from the above ingredients is
\begin{eqnarray}
A^{\rm DT}_{D_{s0}^{*}\bar{D}_s(1^+)} &=&
 c_{K\bar{K},\phi}\,
 c^{1^-}_{\psi\bar{K},D\bar{D}_s}\,
 c_{DK,D_{s0}^*}\, 
 c_{D_{s0}^*\bar{D}_s(1^+)}\,
\nonumber \\
&&\times\! \int\! d^3p_{\bar{D}_s} d^3p_K
{\bm{p}_{\bar{K}K}\cdot \bm{\epsilon}_\phi 
\over W-E_{K}-E_{\bar{K}}-E_{\psi}+i\epsilon}
\nonumber\\
&&\times
{
i(\bm{p}_{\bar{K}\psi}\times \bm{\epsilon}_{\psi})
\cdot \bm{p}_{D\bar{D}_s}\,
\bm{p}_{\bar{D}_s}\cdot\bm{p}_{K_f}
\over W-E_{K}-E_{D}-E_{\bar{D}_s}+i\epsilon}
\nonumber\\
&&\times
{  f_{K\bar{K},\phi}^{1} 
 f_{\psi\bar{K}}^{1} 
 f_{D\bar{D}_s}^{1}
 f_{DK,D_{s0}^*}^{0}
 F_{D_{s0}^*\bar{D}_s K_f,B}^{11}
\over W-E_{D_{s0}^*}-E_{\bar{D}_s}+ {i\over 2} \Gamma_{D^*_{s0}}} ,
\label{eq:dt_ds0ds}
\end{eqnarray}
where the summation over 
$D^{+}K^0\bar{K}^0$ and 
$D^{0}K^+\bar{K}^-$ intermediates states
with the charge dependent particle masses is implicit;
$K^+$ in the final state is denoted by $K_f$, and
$W$ is related to the total energy $E$ by
$W\equiv E-E_{K_f}$.

The $D^*_{s0}$ width ($\Gamma_{D^*_{s0}}$) 
should be small because 
the dominant $D^*_{s0}\to D_s\pi$ decay is isospin-violating.
Experimentally, only an upper limit has been set: $\Gamma_{D^*_{s0}}<3.8$~MeV~\cite{babar_Ds0}.
Theoretically, 
$\Gamma_{D^*_{s0}}\sim 0.1$~MeV (0.01~MeV)
has been given
by a hadron molecule model~\cite{liu_Ds0}
($c\bar{s}$ models~\cite{Godfrey,Colangelo}).
We use $\Gamma_{D^*_{s0}}=0.1$~MeV;
our results do not significantly change
for $\Gamma_{D^*_{s0}}< 1$~MeV.

Similarly,
we consider other
$p$-wave $D_{sJ}^{(*)}\bar{D}_s^{(*)}$ pairs of 
Eq.~(\ref{eq:ddpair}) in DT diagrams.
The $D_{s1}\bar{D}_s$ cusp needs to be $0^+$ to be consistent with 
the LHCb result for $X(4500)$.
Interestingly, 
the DT amplitudes of
the $s$-wave $D_{s1}\bar{D}_s(1^-)$ 
and $p$-wave $D_{s1}\bar{D}_s(1^+)$ 
share the same $D^*\bar{D}_s\to J/\psi\bar{K}$ interaction of
Eq.~(\ref{eq:DD5}), while 
the $p$-wave $D_{s1}\bar{D}_s(0^+)$ DT includes Eq.~(\ref{eq:DD4}).
Thus the dominance of $0^+$ and hindered $1^\pm$
might hint that the $D^*\bar{D}_s\to J/\psi\bar{K}$ interaction of
Eq.~(\ref{eq:DD5}) is weaker than that of Eq.~(\ref{eq:DD4}).
Cusps from
the $D_{s0}^{*}\bar{D}_s^{*}$ and $D_{s1}\bar{D}_s^{*}$ pairs
occur at the dips in the $M_{J/\psi\phi}$
distribution, and the LHCb did not assign any spin-parity to these
structures; we can thus choose their spin-parity to obtain a good fit. 

The $D_s^*\bar{D}_s(1^+)$ one-loop amplitude 
includes 
$B^+\to D_{s}^*\bar{D}_sK^+$
and 
$D_{s}^*\bar{D}_s\to J/\psi\phi$ vertices 
given by
\begin{eqnarray}
&& c_{D_s^*\bar{D}_s(1^+)}\, \bm{p}_K\cdot \bm{\epsilon}_{D_s^*}\,
 F_{D_s^*\bar{D}_s K,B}^{01}\ , \\
&& c^{1^+}_{\psi\phi,D_s^*\bar{D}_s}\,
i(\bm{\epsilon}_\phi\times \bm{\epsilon}_\psi)
\cdot\bm{\epsilon}_{D_s^*}\,
 f_{\psi\phi}^{0}
 f_{D_s^*\bar{D}_s}^{0} \ , 
\end{eqnarray}
respectively, from which 
the one-loop amplitude is
\begin{eqnarray}
A^{\rm 1L}_{D_{s}^{*}\bar{D}_s(1^+)} &=&
c^{1^+}_{\psi\phi,D_s^*\bar{D}_s}\,
c_{D_s^*\bar{D}_s(1^+)}
i(\bm{\epsilon}_\phi\times \bm{\epsilon}_\psi)
\cdot\bm{p}_{K_f} \nonumber\\
&&\times\int d^3p_{\bar{D}_s}
{ 
 f_{\psi\phi}^{0}
 f_{D_s^*\bar{D}_s}^{0}
 F_{D_s^*\bar{D}_s K_f,B}^{01}
\over W-E_{D_s^*}-E_{\bar{D}_s}
+i\epsilon
}\ .
\end{eqnarray}
The $D_s^*$ width 
is expected to be tiny ($\sim 0.1$~keV~\cite{babar_Ds_star,Ds_star_width})
and thus neglected. 

For the $D_s^{*} \bar{D}_s^{(*)}\to J/\psi \phi$ transition
in the one-loop diagram,
we consider a single-channel 
$D_s^{*} \bar{D}_s^{(*)}$ scattering
followed by a perturbative 
$D_s^{*} \bar{D}_s^{(*)}\to J/\psi \phi$ transition.
Details are given in 
Sec.~2 of the Supplemental Material in
Ref.~\cite{DTS}.
Since 
attractive $D_s^{*} \bar{D}_s^{(*)}$ interactions are preferred
in fitting the LHCb data,
we fix the $D_s^{*} \bar{D}_s^{(*)}$ interaction strengths so that
the scattering length ($a$) is 
a moderately attractive value : $a\sim 0.55$~fm~\footnote{
The scattering length $(a)$ is related to the phase shift $(\delta)$ by 
$p\cot\delta=1/a + {\cal O}(p^2)$.}.
This scattering model has a virtual pole at $\sim 20$~MeV below the 
$D_s^{*} \bar{D}_s^{(*)}$ threshold.
Similar virtual poles are also obtained in Ref.~\cite{xkdong2}
where a contact interaction saturated by a $\phi$-exchange mechanism
is used.
An attractive $D_s^{*} \bar{D}_s^{(*)}$ interaction 
makes the threshold cusp significantly sharper~\cite{xkdong}.
Yet, the fit quality does not largely change even when $a=0$ after
adjusting other coupling parameters.

\section{results}

\subsection{Singular behaviors of double triangle amplitudes}

\begin{figure}[t]
\begin{center}
\includegraphics[width=.5\textwidth]{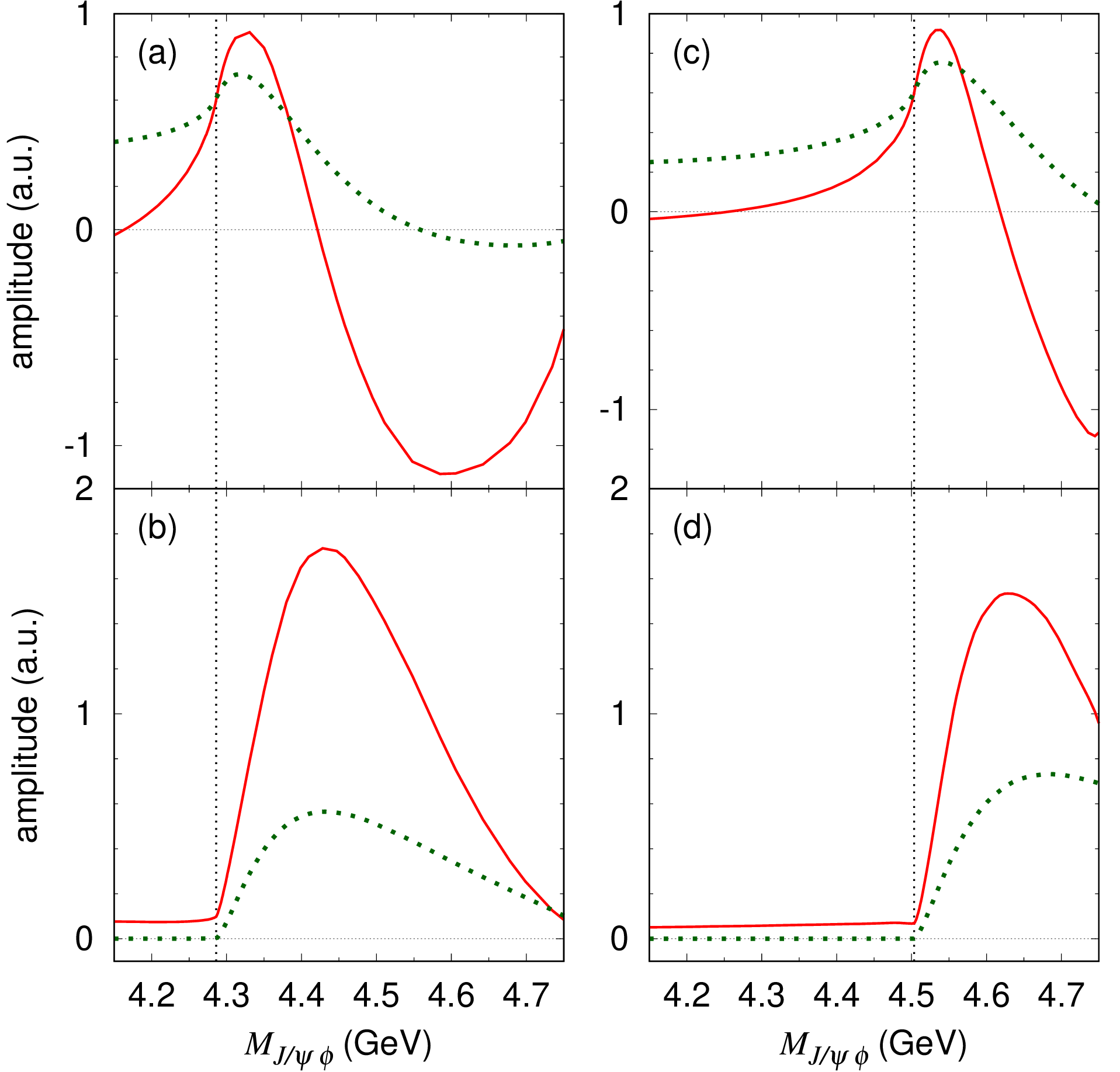}
\end{center}
 \caption{
Double triangle amplitudes; (a) real, and (b) imaginary parts.
The red solid curves
are from Fig.~\ref{fig:diag}(a) with
$D_{sJ}^{(*)}\bar{D}_s^{(*)}(J^P)=D_{s0}^*(2317)^+D_s^-(1^+)$
and $D^{(*)} K\bar{K}= D^+K^0\bar{K}^0 + D^0K^+K^-$.
The green dotted curves are 
a $D_{s0}^*(2317)^+{D}_s^-(1^+)$ one-loop amplitude.
The two amplitudes have been arbitrary scaled to have the same magnitude at the
the $D_{s0}^*(2317)^+{D}_s^-$ threshold
 indicated by the dotted vertical lines.
An overall constant phase factor has been multiplied to the 
double triangle amplitude to compare well with the one-loop amplitude.
The amplitudes in the panel (c) [(d)] are obtained from those in (a)
[(b)] by replacing 
$D_{s0}^*(2317)^+{D}_s^-(1^+)$ 
and $D$ with 
$D_{s1}^*(2536)^+{D}_s^-(0^+)$ 
and $D^*$, respectively.
}
\label{fig:amp}
\end{figure}
\begin{figure}[t]
\begin{center}
\includegraphics[width=.5\textwidth]{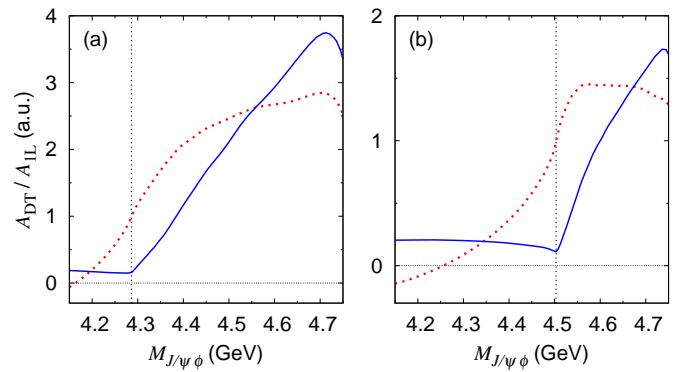}
\end{center}
 \caption{
Ratio of double triangle ($A_{DT}$) and one-loop ($A_{1L}$) amplitudes.
The red dotted and blue solid curves show
${\rm Re}[A_{DT}/A_{1L}]$ and ${\rm Im}[A_{DT}/A_{1L}]$, respectively. 
The ratios in the panels (a) and (b) are obtained using the amplitudes
shown in 
Fig.~\ref{fig:amp}(a,b) and Fig.~\ref{fig:amp}(c,d), respectively.
}
\label{fig:ratio}
\end{figure}
A DT amplitude ($A_{\rm DT}$) such as 
Fig.~\ref{fig:diag}(a) can cause a kinematical singularity (anomalous
threshold)~\cite{s-matrix}.
The DT singularity can generate a resonancelike structure in a decay spectrum,
as first demonstrated in Refs.~\cite{DTS,DTS-pos}.
According to the Coleman-Norton theorem~\cite{coleman}, 
$A_{\rm DT}$ has the {\it leading singularity} 
if the whole DT process is kinematically allowed at the
classical level:
the energy and momentum are always conserved;
in Fig.~\ref{fig:diag}(a),
all internal momenta are collinear
in the $D_{sJ}^{(*)}\bar{D}_{s}^{(*)}$ center-of-mass frame;
$D^{(*)}$ and $\bar{D}_{s}^{(*)}$ 
($\bar{K}$ and $K$) 
are moving to the same direction and the former is faster than the latter.
Whether a given diagram has a singularity is solely
determined by the participating particles' masses.

The DT amplitudes
included in our $B^+\to J/\psi \phi K^+$
model do not cause the leading singularity. 
Yet, the leading singularity is close to being
caused~\footnote{See Appendix~\ref{app2} for a discussion on
how closely $A_{\rm DT}$ satisfies the kinematical condition for 
the leading singularity.
},
and its effect is expected to be visible as
an enhancement of the threshold cusp.
We thus study the singular behavior of 
$A_{\rm DT}$ numerically~\footnote{In
principle, the singular behavior of 
$A_{\rm DT}$ can also be studied more analytically
by examining the corresponding Landau equation~\cite{s-matrix,landau}.}.
In Fig.~\ref{fig:amp}(a,b),
we show $A_{\rm DT}$
of Eq.~(\ref{eq:dt_ds0ds}) by the
red solid curve,
and a one-loop amplitude 
$A_{\rm 1L}$ by the green dotted curve.
The $p$-wave pair of $D_{s0}^*(2317)\bar{D}_s(1^+)$ is included in 
$A_{\rm DT}$ and $A_{\rm 1L}$.
While both amplitudes have threshold cusps at 
the $D_{s0}^*(2317)\bar{D}_s$ threshold, 
$A_{\rm DT}$ 
is sharper.
This can be seen more clearly by taking a ratio 
$A_{\rm DT}/A_{\rm 1L}$ as shown 
in Fig.~\ref{fig:ratio}(a).
The ratio is still singular;
the derivative of the
imaginary part with respect to 
$M_{J/\psi\phi}$ seems divergent
at the threshold.
The ratio may also serve to isolate from $A_{\rm DT}$
the kinematical singularity effect other than 
the ordinary threshold cusp.
Similarly, 
$A_{\rm DT}$ and $A_{\rm 1L}$ including
$D_{s1}(2536)\bar{D}_s(0^+)$ $p$-wave pairs
and their ratio are shown in 
Fig.~\ref{fig:amp}(c,d)
and Fig.~\ref{fig:ratio}(b), respectively.
At the threshold,
$A_{\rm DT}$ is even sharper and
the imaginary part of the ratio is singular. 
The quantitative difference in the singular behavior
 between
$A^{\rm DT}_{D_{s0}^{*}\bar{D}_s(1^+)}$ and 
$A^{\rm DT}_{D_{s1}\bar{D}_s(0^+)}$ is 
from the fact that 
$D_{s1}(2536)\to D^* K$ is allowed at on-shell while 
$D_{s0}^*(2317)\to DK$ is not.

\subsection{Analysis of the LHCb data}

To analyze the $B^+\to J/\psi \phi K^+$ data,
we have seven DT diagrams, four one-loop diagrams, two
$Z_{cs}$-excitation diagrams, and two direct decay diagrams. 
Each of the diagrams 
has a complex overall factor that comes from the product of unknown
coupling constants. 
The overall normalization and phases of the 
$0^+$ and $1^+$ full amplitudes are arbitrary.
We totally have 20 fitting parameters from the coupling constants after 
removing relatively unimportant parameters;
see Tables~\ref{tab:param1} and \ref{tab:para2} in
Appendix~\ref{app1} 
for coupling parameters and fit fractions.
Also, we use cutoffs different from the common value for
the direct decay diagrams so that their
$M_{J/\psi \phi}$ distributions are similar to the phase-space shape.
We note that 
no parameter can adjust the DT and one-loop threshold cusp positions
 where the experimental peaks are located.
Fitting the $M_{J/\psi\phi}$-distribution lineshape requires
the adjustable parameters.
In contrast, the quark and hadron-molecule models need adjustable parameters 
to get pole positions at the experimental peak positions;
parameters for fitting the lineshape are needed additionally.
It is therefore likely that our model can fit the $M_{J/\psi\phi}$-distribution
with fewer parameters.

\begin{figure}[t]
\begin{center}
\includegraphics[width=.5\textwidth]{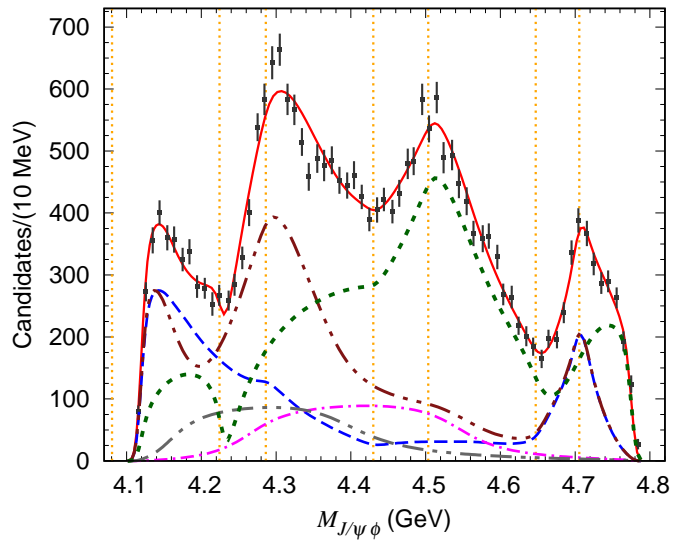}
\end{center}
 \caption{
$J/\psi \phi$ invariant mass $(M_{J/\psi \phi})$ distribution for
$B^+\to J/\psi\phi K^+$.
The red solid 
curve is from the full model.
The $1^+$ and $0^+$ $J/\psi\phi$ partial wave contributions
without [with] $Z_{cs}$ are shown by 
the blue dashed and green dotted 
[brown dash-two-dotted]
curves, respectively;
$Z_{cs}$ is not considered for $0^+$.
The $Z_{cs}(4000)$ [heavier $Z_{cs}$]
contribution alone is given by
the magenta dash-dotted [gray two-dash-two-dotted] curve.
The dotted vertical lines indicate thresholds for,
from left to right, 
$D_{s}^*\bar{D}_s$,
$D_{s}^*\bar{D}^*_s$,
$D_{s0}^*(2317)\bar{D}_s$,
$D_{s0}^*(2317)\bar{D}^*_s$,
$D_{s1}(2536)\bar{D}_s$,
$D_{s1}(2536)\bar{D}^*_s$, and
$\psi'\phi$,  respectively.
Data are from Ref.~\cite{lhcb_phi}.
 }
\label{fig:comp-data}
\end{figure}
We compare in Fig.~\ref{fig:comp-data}
our calculation with the $M_{J/\psi \phi}$ distribution data.
Theoretical curves are smeared with the experimental bin width.
The data are well fitted by the full model (red solid curve).
In particular, 
the resonancelike four peaks and three dips 
are well described by threshold cusps
from the DT and one-loop amplitudes.
We used common cutoff values over $\Lambda=0.8-1.5$~GeV and confirmed 
the stability of the fit quality. 
This is understandable since the structures in the spectrum
are generated by the threshold cusps that are insensitive to
a particular choice of the form factors.

In the same figure, we plot the
$1^+$ partial wave contribution without
$Z_{cs}$ [blue dashed curve].
There are two clear resonancelike 
cusps from
$A^{\rm 1L}_{D_{s}^{*}\bar{D}_s(1^+)}$ and
$A^{\rm 1L}_{\psi'\phi(1^+)}$
at $M_{J/\psi \phi}\sim 4.14$ and 4.7~GeV, respectively.
These threshold cusps would play a role similar to those of 
$X(4140)$ and $X(4685)$ found in the LHCb analysis.
The dip at $M_{J/\psi \phi}\sim 4.65$~GeV is caused by
the $A^{\rm DT}_{D_{s1}\bar{D}^*_s(1^+)}$ cusp.
The $1^+$ contribution also has a relatively small cusp
at $M_{J/\psi \phi}\sim 4.29$~GeV caused by 
$A^{\rm DT}_{D_{s0}^{*}\bar{D}_s(1^+)}$.
This cusp interferes with  
$Z_{cs}(4000)$ accompanied by $p$-wave $\phi$
to create the prominent $X(4274)$ structure 
as seen in the brown dash-two-dotted curve.

Similarly, threshold cusps play a major role to form
resonancelike and dip structures in the 
$0^+$ contribution [green dotted curve].
The cusp from
$A^{\rm DT}_{D_{s1}\bar{D}_s(0^+)}$ 
develops the $X(4500)$ structure.
This structure is made even sharper by the neighboring two dips at 
$M_{J/\psi \phi}\sim 4.46$ and 4.66~GeV 
due to the 
$A^{\rm DT}_{D_{s0}^{*}\bar{D}^*_s(0^+)}$
and
$A^{\rm DT}_{D_{s1}\bar{D}^*_s(0^+)}$
cusps, respectively.
There is another peak at 
$M_{J/\psi \phi}\sim 4.75$~GeV, near the $X(4700)$ peak,
that is caused by the dip at
$M_{J/\psi \phi}\sim 4.66$~GeV 
and the rapidly shrinking phase-space near the kinematical endpoint. 
Another dip is created
at $M_{J/\psi \phi}\sim 4.23$~GeV 
by $A^{\rm 1L}_{D_{s}^{*}\bar{D}^*_s(0^+)}$. 
The contributions from the lighter and heavier $Z_{cs}$
are shown by the magenta dash-dotted and gray two-dash-two-dotted
curves, respectively.
These $Z_{cs}$ contributions without interference
are similar to those of the LHCb analysis~\cite{lhcb_phi}.

The above partial wave decomposition 
might change by including
more data and more partial waves as done in the LHCb amplitude analysis~\cite{lhcb_phi};
this will be a future work
(see also footnote~\ref{foote4}). 
The objective here is to demonstrate that the $X$ and dip structures 
can be well described with the kinematical effects.
We also add that the present analysis by no means excludes
other interpretations for the $X$ states based on 
the quark and hadron molecule models.
We need more experimental and lattice QCD
inputs to judge the different interpretations.

\begin{figure}[t]
\begin{center}
\includegraphics[width=.5\textwidth]{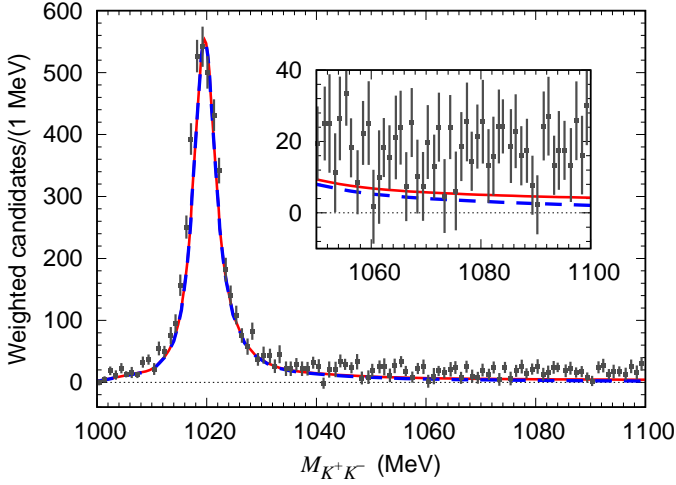}
\end{center}
 \caption{
$K^+K^-$ invariant mass $(M_{K^+K^-})$ distribution for $B^+\to J/\psi K^+K^-K^+$.
The blue dashed curve is from the full model for 
$B^+\to J/\psi\phi K^+$ followed by $\phi\to K^+K^-$;
$M_{K^+K^-}$ is from the $K^+K^-(\leftarrow \phi)$ pair.
The red solid curve additionally includes 
diagrams of Fig.~\ref{fig:diag}(a)
with the $K\bar{K}\to\phi$ vertex removed.
Data are from Ref.~\cite{lhcb_phi2}.
The small window shows the enlarged $\phi$-tail region.
 }
\label{fig:comp-data-kk}
\end{figure}
The LHCb presented 
$M_{K^+K^-}$ distribution for $B^+\to J/\psi K^+K^-K^+$~\cite{lhcb_phi2}.
In Fig.~\ref{fig:comp-data-kk}, the data show the 
$\phi$ peak and a small backgroundlike contribution. 
The data actually put a constraint on the contribution from 
the DT diagrams of Fig.~\ref{fig:diag}(a).
This is because when the DT diagrams followed by $\phi\to K^+K^-$
contribute to $B^+\to J/\psi K^+K^-K^+$, 
there must be a contribution from 
the diagrams of Fig.~\ref{fig:diag}(a) with the last $K\bar{K}\to\phi$
vertex removed.
This single triangle contribution has to be smaller than 
the backgroundlike data. 
Thus, in Fig.~\ref{fig:comp-data-kk}, 
we plot the $M_{K^+K^-}$ distribution from our model
with (red solid curve) and without (blue dashed curve) the single triangle contribution.
The single triangle contribution does not significantly
change the $\phi$ peak and slightly enhances the $\phi$-tail region
well within the experimental constraint. 
The unexplained part of the backgroundlike data should be from
non-$\phi$ mechanisms not
considered here.

The key assumption in our model is that
short-range $D_{sJ}^{(*)}\bar{D}_s^{(*)}\to J/\psi\phi$ transition strengths are
weak. 
The assumption naturally leads to the DT mechanisms that cause enhanced threshold
cusps consistent with the LHCb data. 
If the assumption is wrong, 
the initial $B^+\to D_{sJ}^{(*)}\bar{D}_s^{(*)} K^+$ decays would be followed by
$D_{sJ}^{(*)}\bar{D}_s^{(*)}$ one-loop like Fig.~\ref{fig:diag}(b),
and ordinary threshold cusps are expected.
However, $s$-wave cusps are disfavored by the LHCb data, and 
$p$-wave cusps are too suppressed to fit the data as shown in 
Ref.~\cite{xhliu}.
Thus the LHCb's result seems to be in favor of the assumption.
For $D_{sJ}^{(*)}=D_{s0}^*(2317)$,
the assumption is also partly supported by previous theoretical works that analyzed
lattice QCD energy spectrum and found 
$D_{s0}^*(2317)$ to be mainly a $DK$ molecule~\cite{liu_Ds0,torres_Ds0,Cheung_Ds0,Yang_Ds0}.
On the other hand, 
$D_{s1}(2536)$ have been mostly considered to be a $p$-wave
$c\bar{s}$~\cite{Yang_Ds0,ds1_ortega} and, thus, 
the assumption is not intuitively understandable.
Yet, 
$D_{s1}(2536)$ is known to have a strong coupling to $D^*K$, which has been utilized
in our model. 

\begin{acknowledgments}
I thank F.-K. Guo for useful comments on the manuscript.
This work is in part supported by 
National Natural Science Foundation of China (NSFC) under contracts 
U2032103 and 11625523, 
and also by
National Key Research and Development Program of China under Contracts 2020YFA0406400.
\end{acknowledgments}

\appendix 

\section{$B^+\to J/\psi \phi K^+$ amplitudes}
\label{app1}

We present amplitude formulas for diagrams in Fig.~\ref{fig:diag}.
For double triangle (DT) diagrams of Fig.~\ref{fig:diag}(a),
we consider those including 
$D_{s0}^*\bar{D}_s(1^+)$,
$D_{s0}^*\bar{D}^*_s(0^+,1^+)$,
$D_{s1}\bar{D}_s(0^+,1^+)$, and
$D_{s1}\bar{D}^*_s(0^+,1^+)$
$p$-wave pairs.
Each of the DT diagrams 
includes four vertices.
The initial $B^+\to D_{sJ}^{(*)}\bar{D}^{(*)}_s K^+$ vertices 
are given
by
\begin{eqnarray}
\label{eq:dt11}
&& c_{D_{s0}^*\bar{D}_s(1^+)}\,
\bm{p}_{\bar{D}_s}\cdot\bm{p}_K\,
 F_{D_{s0}^*\bar{D}_s K,B}^{11}\ , \\
&& c_{D_{s0}^*\bar{D}^*_s(0^+)}\,
\bm{p}_{\bar{D}^*_s}\cdot \bm{\epsilon}_{\bar{D}_s^*}\,
 F_{D_{s0}^*\bar{D}^*_s K,B}^{10}\ , \\
&& c_{D_{s0}^*\bar{D}^*_s(1^+)}\,
i(\bm{p}_{\bar{D}^*_s}\times \bm{\epsilon}_{\bar{D}_s^*})\cdot \bm{p}_K\,
 F_{D_{s0}^*\bar{D}^*_s K,B}^{11}\ , \\
&& c_{D_{s1}\bar{D}_s(0^+)}\,
\bm{\epsilon}_{D_{s1}}\cdot \bm{p}_{\bar{D}_s}\,
 F_{D_{s1}\bar{D}_s K,B}^{10}\ , \\
&& c_{D_{s1}\bar{D}_s(1^+)}\,
i(\bm{p}_{\bar{D}_s}\times\bm{\epsilon}_{D_{s1}})\cdot \bm{p}_K\,
 F_{D_{s1}\bar{D}_s K,B}^{11}\ , \\
&& c_{D_{s1}\bar{D}^*_s(0^+)}\,
i(\bm{\epsilon}_{D_{s1}}\times 
\bm{\epsilon}_{\bar{D}^*_{s}})\cdot \bm{p}_{\bar{D}^*_s}\,
 F_{D_{s1}\bar{D}^*_s K,B}^{10}\ , \\
&& c_{D_{s1}\bar{D}^*_s(1^+)}\,
\bm{\epsilon}_{D_{s1}}\cdot\bm{\epsilon}_{\bar{D}^*_{s}}\
\bm{p}_{\bar{D}^*_s}\cdot \bm{p}_{K}\,
 F_{D_{s1}\bar{D}^*_s K,B}^{11}\ , 
\label{eq:dt12}
\end{eqnarray}
respectively,
with complex coupling constants 
$c_{D_{sJ}^{(*)}\bar{D}^{(*)}_s(J^P)}$.
Here and in what follows,
the initial vertices contributing to the final $0^+$ and $1^+$ $J/\psi\phi$ partial waves
are parity-conserving and parity-violating, respectively.
We have used dipole form factors $F_{ijk,l}^{LL'}$ defined by
\begin{eqnarray}
\label{eq:ff1}
 F_{ijk,l}^{LL'} =
 {1\over \sqrt{E_i E_j E_k E_l}}
\left(\frac{\Lambda^2}{\Lambda^2+q_{ij}^2}\right)^{\!\!2+{L\over 2}}\!\!\!
\left(\frac{\Lambda^{\prime 2}}{\Lambda^{\prime 2}+\tilde{p}_k^2}\right)^{\!\!2+{L'\over 2}}\!\!\!\!\!\!\!\!\!\!\!\!\!,
\end{eqnarray}
where $q_{ij}$ ($\tilde{p}_{k}$) is the momentum of $i$ ($k$) in the
$ij$ (total) center-of-mass frame.
The second vertices
$D_{s0}^*\to D K$ and 
$D_{s1}\to D^* K$ are given by
\begin{eqnarray}
 &&c_{DK,D_{s0}^*}\, f_{DK,D_{s0}^*}^{0} \ , \\
 &&c_{D^*K,D_{s1}}\, 
\bm{\epsilon}_{D_{s1}}\cdot \bm{\epsilon}_{D^*}\,
f_{D^*K,D_{s1}}^{0} \ , 
\end{eqnarray}
with form factors $f_{ij,k}^{L}\equiv f_{ij}^{L}/\sqrt{E_k}$ and 
\begin{eqnarray}
 f_{ij}^{L} =
 {1\over \sqrt{E_i E_j}}
\left(\frac{\Lambda^2}{\Lambda^2+q_{ij}^2}\right)^{2+(L/2)}\ .
\label{eq:ff2}
\end{eqnarray}
The third vertices
$D^{(*)}\bar{D}^{(*)}_s\to J/\psi \bar{K}$ 
are $p$-wave interactions
between the pairs with the same spin-parity $j^p=0^-$ or $1^-$,
and are given with coupling constants 
$c^{j^p}_{D^{(*)}\bar{D}^{(*)}_s,\psi\bar{K}}$
as
\begin{eqnarray}
 \label{eq:DD1}
&& c^{1^-}_{\psi\bar{K},D\bar{D}_s}\,
i(\bm{p}_{\bar{K}\psi}\times \bm{\epsilon}_{\psi})\cdot 
\bm{p}_{D\bar{D}_s}\,
 f_{\psi\bar{K}}^{1} 
 f_{D\bar{D}_s}^{1}, 
\\
 \label{eq:DD2}
&& c^{0^-}_{\psi\bar{K},D\bar{D}^*_s}\,
\bm{p}_{D\bar{D}^*_s}\cdot\bm{\epsilon}_{\bar{D}^*_s}\,
\bm{p}_{\bar{K}\psi}\cdot\bm{\epsilon}_{\psi}\,
 f_{\psi\bar{K}}^{1} 
 f_{D\bar{D}^*_s}^{1}, \\
 \label{eq:DD3}
&& c^{1^-}_{\psi\bar{K},D\bar{D}^*_s}\,
(\bm{p}_{D\bar{D}^*_s}\!\times\!\bm{\epsilon}_{\bar{D}^*_s})\cdot
(\bm{p}_{\bar{K}\psi}\!\times\!\bm{\epsilon}_{\psi})\,
f_{\psi\bar{K}}^{1} f_{D\bar{D}^*_s}^{1} \ , 
%\\
\end{eqnarray}
\begin{eqnarray}
 \label{eq:DD4}
&& c^{0^-}_{\psi\bar{K},D^*\bar{D}_s}\,
\bm{p}_{D^*\bar{D}_s}\!\cdot\!\bm{\epsilon}_{D^*}\,
\bm{p}_{\bar{K}\psi}\!\cdot\!\bm{\epsilon}_{\psi}\,
 f_{\psi\bar{K}}^{1}
 f_{D^*\bar{D}_s}^{1}\ , \\
 \label{eq:DD5}
&& c^{1^-}_{\psi\bar{K},D^*\bar{D}_s}\,
(\bm{p}_{D^*\bar{D}_s}\!\times\!\bm{\epsilon}_{D^*})\cdot
(\bm{p}_{\bar{K}\psi}\!\times\!\bm{\epsilon}_{\psi}) \,
f_{\psi\bar{K}}^{1} f_{D^*\bar{D}_s}^{1}\ , 
\\
 \label{eq:DD6}
&& c^{0^-}_{\psi\bar{K},D^*\bar{D}^*_s}\,
i (\bm{\epsilon}_{D^*}\!\times\!\bm{\epsilon}_{\bar{D}^*_s})
\!\cdot\!\bm{p}_{D^*\bar{D}^*_s}\,
\bm{p}_{\bar{K}\psi}\!\cdot\!\bm{\epsilon}_{\psi}\,
 f_{\psi\bar{K}}^{1}
 f_{D^*\bar{D}^*_s}^{1} , \\
 \label{eq:DD7}
&& c^{1^-}_{\psi\bar{K},D^*\bar{D}^*_s}\,
\bm{\epsilon}_{D^*}\!\cdot\!\bm{\epsilon}_{\bar{D}^*_s}\,
i(\bm{p}_{\bar{K}\psi}\!\times\!\bm{\epsilon}_{\psi})\!
\cdot\!\bm{p}_{D^*\bar{D}^*_s}\,
 f_{\psi\bar{K}}^{1} f_{D^*\bar{D}^*_s}^{1} , \\\nonumber
\end{eqnarray}
respectively,
where a notation of $\bm{p}_{ab}\equiv \bm{p}_a-\bm{p}_b$ has been used.
The fourth vertex $K\bar{K}\to\phi$ is
common for all the DT diagram, and is given as
\begin{eqnarray}
&& c_{K\bar{K},\phi}\,
\bm{p}_{\bar{K}K}
\cdot \bm{\epsilon}_\phi \,
 f_{K\bar{K},\phi}^{1} \ .
\end{eqnarray}
We denote the DT amplitudes
including $D_{sJ}^{(*)}\bar{D}^{(*)}_s(J^P$)
by $A^{\rm DT}_{D_{sJ}^{(*)}\bar{D}^{(*)}_s(J^P)}$.
The DT amplitudes are constructed with the above ingredients,
and are given by
\begin{widetext}

\begin{eqnarray}
\label{eq:DT1}
A^{\rm DT}_{D_{s0}^{*}\bar{D}_s(1^+)} &=&
 c_{K\bar{K},\phi}\,
 c^{1^-}_{\psi\bar{K},D\bar{D}_s}\,
 c_{DK,D_{s0}^*}\, 
 c_{D_{s0}^*\bar{D}_s(1^+)}\,
\int d^3p_{\bar{D}_s}\int d^3p_K
{\bm{p}_{\bar{K}K}\cdot \bm{\epsilon}_\phi 
\over W-E_{K}-E_{\bar{K}}-E_{\psi}+i\epsilon}
\nonumber\\
&&\times
{
i(\bm{p}_{\bar{K}\psi}\times \bm{\epsilon}_{\psi})
\cdot \bm{p}_{D\bar{D}_s}\,
\bm{p}_{\bar{D}_s}\cdot\bm{p}_{K_f}
\over W-E_{K}-E_{D}-E_{\bar{D}_s}+i\epsilon}
{  f_{K\bar{K},\phi}^{1} 
 f_{\psi\bar{K}}^{1} 
 f_{D\bar{D}_s}^{1}
 f_{DK,D_{s0}^*}^{0}
 F_{D_{s0}^*\bar{D}_s K_f,B}^{11}
\over W-E_{D_{s0}^*}-E_{\bar{D}_s}+ {i\over 2} \Gamma_{D^*_{s0}}}\ ,
\\
\label{eq:DT2}
A^{\rm DT}_{D_{s0}^{*}\bar{D}^*_s(0^+)} &=&
 c_{K\bar{K},\phi}\,
 c^{0^-}_{\psi\bar{K},D\bar{D}^*_s}\,
 c_{DK,D_{s0}^*}\,
 c_{D_{s0}^*\bar{D}^*_s(0^+)}\,
\int d^3p_{\bar{D}^*_s}\int d^3p_K
{
\bm{p}_{\bar{K}K}\cdot \bm{\epsilon}_\phi\,
\over W-E_{K}-E_{\bar{K}}-E_{\psi}+i\epsilon}
\nonumber\\
&&\times
{
\bm{p}_{\bar{K}\psi}\cdot\bm{\epsilon}_{\psi}\,
\bm{p}_{D\bar{D}^*_s}\cdot\bm{p}_{\bar{D}^*_s}
\over W-E_{K}-E_{D}-E_{\bar{D}^*_s}+i\epsilon}
{  f_{K\bar{K},\phi}^{1} 
 f_{\psi\bar{K}}^{1} 
 f_{D\bar{D}^*_s}^{1}
 f_{DK,D_{s0}^*}^{0}
 F_{D_{s0}^*\bar{D}^*_s K_f,B}^{10}
\over W-E_{D_{s0}^*}-E_{\bar{D}^*_s}+ {i\over 2} \Gamma_{D^*_{s0}}}\ ,
\\
\label{eq:DT3}
A^{\rm DT}_{D_{s0}^{*}\bar{D}^*_s(1^+)} &=&
 c_{K\bar{K},\phi}\,
c^{1^-}_{\psi\bar{K},D\bar{D}^*_s}\,
c_{DK,D_{s0}^*}\, 
c_{D_{s0}^*\bar{D}^*_s(1^+)}\,
\int d^3p_{\bar{D}^*_s}\int d^3p_K
{
\bm{p}_{\bar{K}K}\cdot \bm{\epsilon}_\phi\,
\over W-E_{K}-E_{\bar{K}}-E_{\psi}+i\epsilon}
\nonumber\\
&&\times
{
i(\bm{p}_{\bar{K}\psi}\times\bm{\epsilon}_{\psi})\cdot
[\bm{p}_{D\bar{D}^*_s}\times (\bm{p}_{K_f}\times\bm{p}_{\bar{D}^*_s})]
\over W-E_{K}-E_{D}-E_{\bar{D}^*_s}+i\epsilon}
{  f_{K\bar{K},\phi}^{1} 
f_{\psi\bar{K}}^{1} 
f_{D\bar{D}^*_s}^{1}
f_{DK,D_{s0}^*}^{0}
F_{D_{s0}^*\bar{D}^*_s K_f,B}^{11}
\over W-E_{D_{s0}^*}-E_{\bar{D}^*_s}+ {i\over 2} \Gamma_{D^*_{s0}} }\ ,
\\
\label{eq:DT4}
A^{\rm DT}_{D_{s1}\bar{D}_s(0^+)} &=&
 c_{K\bar{K},\phi}\,
 c^{0^-}_{\psi\bar{K},D^*\bar{D}_s}\,
 c_{D^*K,D_{s1}}\,
 c_{D_{s1}\bar{D}_s(0^+)}\,
\int d^3p_{\bar{D}_s}\int d^3p_K
{
\bm{p}_{\bar{K}K}\cdot \bm{\epsilon}_\phi\,
\over W-E_{K}-E_{\bar{K}}-E_{\psi}+i\epsilon}
\nonumber\\
&&\times
{
\bm{p}_{\bar{K}\psi}\cdot\bm{\epsilon}_{\psi}\,
\bm{p}_{D^*\bar{D}_s}\cdot\bm{p}_{\bar{D}_s}
\over W-E_{K}-E_{D^*}-E_{\bar{D}_s}+i\epsilon}
{  f_{K\bar{K},\phi}^{1} 
 f_{\psi\bar{K}}^{1}
 f_{D^*\bar{D}_s}^{1}
 f_{D^*K,D_{s1}}^{0}
 F_{D_{s1}\bar{D}_s K_f,B}^{10}
\over W-E_{D_{s1}}-E_{\bar{D}_s}+ {i\over 2} \Gamma_{D_{s1}}}\ ,
\\
\label{eq:DT5}
A^{\rm DT}_{D_{s1}\bar{D}_s(1^+)} &=&
 c_{K\bar{K},\phi}\,
 c^{1^-}_{\psi\bar{K},D^*\bar{D}_s}\,
 c_{D^*K,D_{s1}}\,
 c_{D_{s1}\bar{D}_s(1^+)}\,
\int d^3p_{\bar{D}_s}\int d^3p_K
{
\bm{p}_{\bar{K}K}\cdot \bm{\epsilon}_\phi\,
\over W-E_{K}-E_{\bar{K}}-E_{\psi}+i\epsilon}
\nonumber\\
&&\times
{
i(\bm{p}_{\bar{K}\psi}\times\bm{\epsilon}_{\psi})\cdot
[\bm{p}_{D^*\bar{D}_s}\times (\bm{p}_{K_f}\times\bm{p}_{\bar{D}_s})]
\over W-E_{K}-E_{D^*}-E_{\bar{D}_s}+i\epsilon}
{  f_{K\bar{K},\phi}^{1} 
 f_{\psi\bar{K}}^{1}
 f_{D^*\bar{D}_s}^{1}
 f_{D^*K,D_{s1}}^{0}
 F_{D_{s1}\bar{D}_s K_f,B}^{11}
\over W-E_{D_{s1}}-E_{\bar{D}_s}+ {i\over 2} \Gamma_{D_{s1}}}\ ,
\\
\label{eq:DT6}
A^{\rm DT}_{D_{s1}\bar{D}^*_s(0^+)} &=&
 -2\, c_{K\bar{K},\phi}\,
 c^{0^-}_{\psi\bar{K},D^*\bar{D}^*_s}\,
 c_{D^*K,D_{s1}}\, 
 c_{D_{s1}\bar{D}^*_s(0^+)}\,
\int d^3p_{\bar{D}^*_s}\int d^3p_K
{
\bm{p}_{\bar{K}K}\cdot \bm{\epsilon}_\phi
\over W-E_{K}-E_{\bar{K}}-E_{\psi}+i\epsilon}
\nonumber\\
&&\times
{
\bm{p}_{\bar{K}\psi}\cdot\bm{\epsilon}_{\psi} \,
\bm{p}_{D^*\bar{D}^*_s}\cdot\bm{p}_{\bar{D}^*_s}
\over W-E_{K}-E_{D^*}-E_{\bar{D}^*_s}+i\epsilon}
{  f_{K\bar{K},\phi}^{1} 
 f_{\psi\bar{K}}^{1}
 f_{D^*\bar{D}^*_s}^{1}
 f_{D^*K,D_{s1}}^{0}
 F_{D_{s1}\bar{D}^*_s K_f,B}^{10}
\over W-E_{D_{s1}}-E_{\bar{D}^*_s}+ {i\over 2} \Gamma_{D_{s1}}}\ ,
\\
\label{eq:DT7}
A^{\rm DT}_{D_{s1}\bar{D}^*_s(1^+)} &=&
 3\, c_{K\bar{K},\phi}\,
 c^{1^-}_{\psi\bar{K},D^*\bar{D}^*_s}\,
 c_{D^*K,D_{s1}}\, 
 c_{D_{s1}\bar{D}^*_s(1^+)}\,
\int d^3p_{\bar{D}^*_s}\int d^3p_K
{
\bm{p}_{\bar{K}K}\cdot \bm{\epsilon}_\phi\,
\over W-E_{K}-E_{\bar{K}}-E_{\psi}+i\epsilon}
\nonumber\\
&&\times
{
i(\bm{p}_{\bar{K}\psi}\times\bm{\epsilon}_{\psi})
\cdot\bm{p}_{D^*\bar{D}^*_s}\,
\bm{p}_{\bar{D}^*_s}\cdot \bm{p}_{K_f}\,
\over W-E_{K}-E_{D^*}-E_{\bar{D}^*_s}+i\epsilon}
{  f_{K\bar{K},\phi}^{1} 
 f_{\psi\bar{K}}^{1} 
 f_{D^*\bar{D}^*_s}^{1}
 f_{D^*K,D_{s1}}^{0}
 F_{D_{s1}\bar{D}^*_s K_f,B}^{11}
\over W-E_{D_{s1}}-E_{\bar{D}^*_s}+ {i\over 2} \Gamma_{D_{s1}}}\ ,
\end{eqnarray}
\end{widetext}
where, in each amplitude, the summation over 
$D^{(*)+}K^0\bar{K}^0$ and 
$D^{(*)0}K^+\bar{K}^-$ intermediates states
with the charge dependent particle masses is implicit;
$K^+$ in the final state is denoted by $K_f$, and
$W\equiv E-E_{K_f}$.
Regarding the $D^{(*)}_{sJ}$ widths in the third energy denominators,
while $\Gamma_{D_{s1}}$ is well determined experimentally, 
$\Gamma_{D^*_{s0}}$ is given only an upper limit~\cite{pdg}.
We use $\Gamma_{D^*_{s0}}=0.1$~MeV;
the result does not significantly change for $\Gamma_{D^*_{s0}}<1$~MeV.
We neglect $\Gamma_{D^{*}}$ and $\Gamma_{D^*_{s}}$ which are expected
to be very small
($\Gamma_{D^{*}}\sim 55$~keV~\cite{sxn_x},
$\Gamma_{D^*_{s}}\sim 0.1$~keV~\cite{babar_Ds_star,Ds_star_width}).
From Eqs.~(\ref{eq:DT1})-(\ref{eq:DT7}), we remove terms including 
$\bm{p}_{\phi}\cdot \bm{\epsilon}_\phi$ and $\bm{p}_{\psi}\cdot \bm{\epsilon}_\psi$,
to maintain a consistency with 
the Lorentz condition,
$p_{\phi}\cdot \epsilon_\phi= p_{\psi}\cdot \epsilon_\psi=0$, 
of a relativistic formulation.

For a given DT amplitude, we can analytically integrate
the angular part of the loop-integrals by ignoring smaller angle dependences
from denominators and form factors.
If the DT amplitude is (non-)vanishing after this angular integral, 
the DT integrand has a suppressed (favored) tensor structure.
The $p$-wave $D^{(*)}\bar{D}^{(*)}_s\to J/\psi \bar{K}$ 
interactions of Eqs.~(\ref{eq:DD1})-(\ref{eq:DD7})
are chosen so that 
the DT integrands of Eqs.~(\ref{eq:DT1})-(\ref{eq:DT7}) have
favored tensor structures.
We did not use $s$-wave $D^{(*)}\bar{D}^{(*)}_s\to J/\psi \bar{K}$ 
interactions because the resultant DT integrands have suppressed tensor
structures. 
We numerically confirmed the suppression.

Next we present formulas for 
one-loop amplitudes of
Fig.~\ref{fig:diag}(b) 
including 
$s$-wave pairs of 
$D_s^*\bar{D}_s(1^+)$,
$D_s^*\bar{D}^*_s(0^+)$,
and
$\psi'\phi(0^+,1^+)$
in the loop.
The one-loop processes are initiated by 
$B^+\to D_s^{*}\bar{D}^{(*)}_s K^+$ 
and 
$B^+\to \psi'\phi K^+$ vertices
given as
\begin{eqnarray}
\label{eq:a27}
&& c_{D_s^*\bar{D}_s(1^+)}\, \bm{p}_K\cdot \bm{\epsilon}_{D_s^*}\,
 F_{D_s^*\bar{D}_s K,B}^{01}\ , \\
&& c_{D_s^*\bar{D}_s^*(0^+)}\,
\bm{\epsilon}_{{D}_s^*} \cdot \bm{\epsilon}_{\bar{D}_s^*}\,
 F_{D_s^*\bar{D}_s^* K,B}^{00}\ , \\
&& c_{\psi'\phi(0^+)}\,
\bm{\epsilon}^\prime_{\phi}\cdot\bm{\epsilon}_{\psi'}  \,
 F_{\psi'\phi K,B}^{00}\ , \\
&& c_{\psi'\phi(1^+)}\,
i(\bm{\epsilon}^\prime_{\phi}\times \bm{\epsilon}_{\psi'})\cdot\bm{p}_K\,
 F_{\psi'\phi K,B}^{01}\ . \\\nonumber
\label{eq:dt14}
\end{eqnarray}
The subsequent
$D_{s}^{*}\bar{D}^{(*)}_s, \psi'\phi\to J/\psi\phi$ interactions in $J^P$ partial waves
are given, with coupling constants 
$c^{J^P}_{\psi\phi,D_{s}^{*}\bar{D}^{(*)}_s}$
and $c^{J^P}_{\psi\phi,\psi'\phi}$, as
\begin{eqnarray}
&& c^{1^+}_{\psi\phi,D_s^*\bar{D}_s}\,
i(\bm{\epsilon}_\phi\times \bm{\epsilon}_\psi)
\cdot\bm{\epsilon}_{D_s^*}\,
 f_{\psi\phi}^{0}
 f_{D_s^*\bar{D}_s}^{0} \ , \\
&& c^{0^+}_{\psi\phi,D_s^*\bar{D}^*_s}\,
\bm{\epsilon}_\phi\cdot \bm{\epsilon}_\psi\,
\bm{\epsilon}_{D_s^*}\cdot\bm{\epsilon}_{\bar{D}_s^*}\,
 f_{\psi\phi}^{0}
 f_{D_s^*\bar{D}^*_s}^{0} \ , \\
&& c^{0^+}_{\psi\phi,\psi'\phi}\,
\bm{\epsilon}_\phi\cdot \bm{\epsilon}_\psi\,
\bm{\epsilon}^\prime_{\phi}\cdot\bm{\epsilon}_{\psi'}\,
 f_{\psi\phi}^{0}
 f_{\psi'\phi}^{0} \ , \\
&& c^{1^+}_{\psi\phi,\psi'\phi}\,
(\bm{\epsilon}_\phi\times \bm{\epsilon}_\psi)\cdot
(\bm{\epsilon}^\prime_{\phi}\times\bm{\epsilon}_{\psi'})\,
 f_{\psi\phi}^{0}
 f_{\psi'\phi}^{0} \ .
\end{eqnarray}
We denote 
the one-loop amplitudes 
including $D_{s}^{*}\bar{D}^{(*)}_s (J^P)$
and $\psi'\phi (J^P)$
by $A^{\rm 1L}_{D_{s}^{*}\bar{D}^{(*)}_s (J^P)}$
and $A^{\rm 1L}_{\psi'\phi (J^P)}$, respectively.
The amplitudes are given with the above ingredients as
\begin{widetext}
\begin{eqnarray}
\label{eq:1L1}
A^{\rm 1L}_{D_{s}^{*}\bar{D}_s(1^+)} &=&
c^{1^+}_{\psi\phi,D_s^*\bar{D}_s}\,
c_{D_s^*\bar{D}_s(1^+)}
i(\bm{\epsilon}_\phi\times \bm{\epsilon}_\psi)
\cdot\bm{p}_{K_f}
\int d^3p_{\bar{D}_s}
{ 
 f_{\psi\phi}^{0}
 f_{D_s^*\bar{D}_s}^{0}
 F_{D_s^*\bar{D}_s K_f,B}^{01}
\over W-E_{D_s^*}-E_{\bar{D}_s}
+i\epsilon
}\ ,
\\
\label{eq:1L2}
A^{\rm 1L}_{D_{s}^{*}\bar{D}^*_s(0^+)} &=& 
 3\,c^{0^+}_{\psi\phi,D_s^*\bar{D}_s^*}\,
 c_{D_s^*\bar{D}_s^*(0^+)}\,
\bm{\epsilon}_\phi\cdot \bm{\epsilon}_\psi
\int d^3p_{\bar{D}^*_s}
{   
 f_{\psi\phi}^{0}
 f_{D_s^*\bar{D}^*_s}^{0} 
 F_{D_s^*\bar{D}_s^* K_f,B}^{00}
\over W-E_{D_s^*}-E_{\bar{D}^*_s}
+i\epsilon
}\ ,
\\
\label{eq:1L3}
A^{\rm 1L}_{\psi'\phi(0^+)} &=& 
 3\,c^{0^+}_{\psi\phi,\psi'\phi}\,
 c_{\psi'\phi(0^+)}\,
\bm{\epsilon}_\phi\cdot \bm{\epsilon}_\psi
\int d^3p_{\psi'}
{   
 f_{\psi\phi}^{0}
 f_{\psi'\phi}^{0} 
 F_{\psi'\phi K_f,B}^{00}
\over W-E_{\psi'}-E_{\phi}
+ {i\over 2}\Gamma_{\phi}
}\ ,
\\
\label{eq:1L4}
A^{\rm 1L}_{\psi'\phi(1^+)} &=& 
2\, c^{1^+}_{\psi\phi,\psi'\phi}\,
 c_{\psi'\phi(1^+)}\,
i (\bm{\epsilon}_\phi\times \bm{\epsilon}_\psi)\cdot\bm{p}_{K_f}
\int d^3p_{\psi'}
{   
 f_{\psi\phi}^{0}
 f_{\psi'\phi}^{0}
 F_{\psi'\phi K_f,B}^{01}
\over W-E_{\psi'}-E_{\phi}
+ {i\over 2}\Gamma_{\phi}
}\ ,
\end{eqnarray}
\end{widetext}
where $\Gamma_{\psi'}$ has been neglected since 
$\Gamma_{\psi'}\ll \Gamma_{\phi}$.

\begin{table*}[t]
\renewcommand{\arraystretch}{1.6}
\tabcolsep=3.mm
\caption{\label{tab:param1} Fit fractions and parameter values.
The common cutoff value $\Lambda=1$~GeV is used.
The first column lists each mechanism considered in our model,
 and the second column is its fit fraction (\%)
defined in Eq.~(\ref{eq:ff}).
The third column lists the product of coupling constants to fit the data, 
and its value and unit are given in the fourth and fifth columns, respectively.
Amplitude formulas are given in the equations in 
the last column.
}

\begin{tabular}{lrlclr}
$A^{\rm DT}_{D_{s0}^{*}\bar{D}_s(1^+)}$   &  17.5&$c_{K\bar{K},\phi}\, c^{1^-}_{\psi\bar{K},D\bar{D}_s}\,   c_{DK,D_{s0}^*}\,   c_{D_{s0}^*\bar{D}_s(1^+)}$ &$  -158.-57.   \,i$& GeV$^{-3}$& Eq.~(\ref{eq:DT1})\\
$A^{\rm DT}_{D_{s0}^{*}\bar{D}^*_s(0^+)}$ &   5.7&$c_{K\bar{K},\phi}\, c^{0^-}_{\psi\bar{K},D\bar{D}^*_s}\, c_{DK,D_{s0}^*}\, c_{D_{s0}^*\bar{D}^*_s(0^+)}$ &$  44.6           $& GeV$^{-2}$& Eq.~(\ref{eq:DT2})\\
$A^{\rm DT}_{D_{s0}^{*}\bar{D}^*_s(1^+)}$ &   2.4&$c_{K\bar{K},\phi}\, c^{1^-}_{\psi\bar{K},D\bar{D}^*_s}\, c_{DK,D_{s0}^*}\, c_{D_{s0}^*\bar{D}^*_s(1^+)}$ &$ -32.0 +16.0  \,i$& GeV$^{-3}$& Eq.~(\ref{eq:DT3})\\
$A^{\rm DT}_{D_{s1}\bar{D}_s(0^+)}$       &   6.4&$c_{K\bar{K},\phi}\, c^{0^-}_{\psi\bar{K},D^*\bar{D}_s}\, c_{D^*K,D_{s1}}\,     c_{D_{s1}\bar{D}_s(0^+)}$ &$ -45.2           $& GeV$^{-2}$& Eq.~(\ref{eq:DT4})\\
$A^{\rm DT}_{D_{s1}\bar{D}_s(1^+)}$       &   -- &$c_{K\bar{K},\phi}\, c^{1^-}_{\psi\bar{K},D^*\bar{D}_s}\, c_{D^*K,D_{s1}}\,     c_{D_{s1}\bar{D}_s(1^+)}$ & 0 (fixed)	        & GeV$^{-3}$& Eq.~(\ref{eq:DT5})\\
$A^{\rm DT}_{D_{s1}\bar{D}^*_s(0^+)}$     &   3.9&$c_{K\bar{K},\phi}\, c^{0^-}_{\psi\bar{K},D^*\bar{D}^*_s}\, c_{D^*K,D_{s1}}\, c_{D_{s1}\bar{D}^*_s(0^+)}$ &$  24.0 -15.6  \,i$& GeV$^{-2}$& Eq.~(\ref{eq:DT6})\\
$A^{\rm DT}_{D_{s1}\bar{D}^*_s(1^+)}$     &   4.8&$c_{K\bar{K},\phi}\, c^{1^-}_{\psi\bar{K},D^*\bar{D}^*_s}\, c_{D^*K,D_{s1}}\, c_{D_{s1}\bar{D}^*_s(1^+)}$ &$ -74.4           $& GeV$^{-3}$& Eq.~(\ref{eq:DT7})\\
%					   %
$A^{\rm 1L}_{D_{s}^{*}\bar{D}_s(1^+)}$    &  24.5&$c^{1^+}_{\psi\phi,D_s^*\bar{D}_s}\, c_{D_s^*\bar{D}_s(1^+)}$      & $ 85.5           $& GeV$^{-1}$&  Eq.~(\ref{eq:1L1})\\
$A^{\rm 1L}_{D_{s}^{*}\bar{D}^*_s(0^+)}$  &   5.8&$c^{0^+}_{\psi\phi,D_s^*\bar{D}_s^*}\, c_{D_s^*\bar{D}_s^*(0^+)}$  & $ -4.89 +3.63 \,i$&        $\quad -$&  Eq.~(\ref{eq:1L2})\\
$A^{\rm 1L}_{\psi'\phi(0^+)}$             &   -- &$c^{0^+}_{\psi\phi,\psi'\phi}\, c_{\psi'\phi(0^+)}$                & $    0$ (fixed)   &        $\quad -$&  Eq.~(\ref{eq:1L3})\\
$A^{\rm 1L}_{\psi'\phi(1^+)}$             &  36.1&$c^{1^+}_{\psi\phi,\psi'\phi}\, c_{\psi'\phi(1^+)}$                & $138.  -54.   \,i$& GeV$^{-1}$&  Eq.~(\ref{eq:1L4})\\
%					   %
$A^{0^+}_{Z_{cs}}$                        &   -- &$c^{0^+}_{Z_{cs}} $ &$    0$ (fixed)        & GeV$^2$& Eq.~(\ref{eq:Zcs1})\\
$A^{1^+}_{Z_{cs}}$                        &  10.7&$c^{1^+}_{Z_{cs}} $ &$  -11.4    -37.2  \,i$& GeV    & Eq.~(\ref{eq:Zcs2})\\
$A^{0^+}_{Z_{cs}^{\prime}}$               &   -- &$c^{0^+}_{Z'_{cs}}$ &$    0$ (fixed)        & GeV$^2$& Eq.~(\ref{eq:Zcs1})\\
$A^{1^+}_{Z_{cs}^{\prime}}$               &   9.4&$c^{1^+}_{Z'_{cs}}$ &$   27.4    +38.0  \,i$& GeV    & Eq.~(\ref{eq:Zcs2})\\
%					   %
$A_{\rm dir}^{0^+}$                       & 66.8&$c_{\rm dir}^{0^+}$ &$-134.$&        $\quad -$& Eq.~(\ref{eq:dir_s0})\\
$A_{\rm dir}^{1^+}$                       & 45.1&$c_{\rm dir}^{1^+}$ &$ 445.$& GeV$^{-1}$& Eq.~(\ref{eq:dir_s1})\\
\end{tabular}
\end{table*}

\begin{table}[t]
\renewcommand{\arraystretch}{1.5}
\tabcolsep=4.5mm
\caption{\label{tab:para2}
Parameter values for the full model ($\Lambda=1$~GeV)
{\it not} fitted to the LHCb data~\cite{lhcb_phi}.
The last two parameters are elastic $D_s^*\bar{D}^{(*)}_s$
interaction strengths defined by
Eq.~(43) in the Supplemental Material of Ref.~\cite{DTS}.
}
\begin{tabular}[b]{lrr}
$\Lambda$ (MeV) &1000 &  \\
$\Lambda_{\rm dir}^{^\prime 0^+}$ (MeV) &850 &Eq.~(\ref{eq:dir_s0})  \\
$\Lambda_{\rm dir}^{^\prime 1^+}$ (MeV) &630 &Eq.~(\ref{eq:dir_s1})  \\
$h_{D_s^*\bar{D}_s(1^+)}$   & $-2$ & \\
$h_{D_s^*\bar{D}^*_s(0^+)}$ & $-2$ & 
\end{tabular}
\end{table}

Regarding the $Z_{cs}$ excitation mechanisms [Fig.~\ref{fig:diag}(c)],
we consider lighter and heavier ones, respectively denoted by
$Z_{cs}$ and $Z_{cs}'$,
that could be identified with
$Z_{cs}(4000)$ and $Z_{cs}(4220)$ from the LHCb analysis~\cite{lhcb_phi}.
Our $Z_{cs}^{(\prime)}$ would simulate
the $D_s^{(*)+}\bar{D}^{*0}$ threshold cusps
enhanced by virtual states found in 
coupled-channel analyses~\cite{zcs_model1,zcs_model2,zcs_model3}.
We consider $s$- and $p$-wave $B^+\to Z_{cs}^{(\prime)+}\phi$ decays
followed by $Z_{cs}^{(\prime)+}\to J/\psi K^+$. 
The $s$- and $p$-wave decays are parity-conserving and parity-violating,
respectively, 
and contribute to the $0^+$ and $1^+$
$J/\psi\phi$ final states, respectively.
The corresponding amplitudes $A^{J^P}_{Z_{cs}^{(\prime)}}$
are given by
\begin{eqnarray}
\label{eq:Zcs1}
 A^{0^+}_{Z_{cs}^{(\prime)}} &=&
c^{0^+}_{Z_{cs}^{(\prime)}}\,
{\bm{\epsilon}_\phi\,\cdot \bm{\epsilon}_\psi\,
f_{\psi K_f,Z_{cs}^{(\prime)}}^{0}
f_{Z_{cs}^{(\prime)}\phi,B}^{0}
\over E-E_{\phi}-E_{Z_{cs}^{(\prime)}}+{i\over 2}
(\Gamma_{Z_{cs}^{(\prime)}}+\Gamma_{\phi})},
 \\
\label{eq:Zcs2}
 A^{1^+}_{Z_{cs}^{(\prime)}} &=&
c^{1^+}_{Z_{cs}^{(\prime)}}\,
{i(\bm{\epsilon}_\phi\times \bm{\epsilon}_\psi)\cdot \bm{p}_\phi\,
f_{\psi K_f,Z_{cs}^{(\prime)}}^{0}
f_{Z_{cs}^{(\prime)}\phi,B}^{1}
\over E-E_{\phi}-E_{Z_{cs}^{(\prime)}}+{i\over 2}
(\Gamma_{Z_{cs}^{(\prime)}}+\Gamma_{\phi})} ,
\end{eqnarray}
where the $Z_{cs}$ and $Z_{cs}^{\prime}$ masses are 
3975~MeV and 4119~MeV from 
the $D_s^+\bar{D}^{*0}$ and
$D_s^{*+}\bar{D}^{*0}$ thresholds, respectively; 
their widths are set to be 100~MeV (constants).

The direct decay 
amplitudes [Fig.~\ref{fig:diag}(d)] can be projected onto the
$J/\psi\phi(J^P)$ partial waves.
Thus we employ a form 
as follows:
\begin{eqnarray}
\label{eq:dir_s0}
 A_{\rm dir}^{0^+} &=&
  c_{\rm dir}^{0^+}\,
\bm{\epsilon}_\phi \cdot \bm{\epsilon}_\psi  \,
 F_{\psi \phi K_f,B}^{00} \ ,
\\
\label{eq:dir_s1}
 A_{\rm dir}^{1^+} &=&
c_{\rm dir}^{1^+}\, 
i(\bm{\epsilon}_\phi \times \bm{\epsilon}_\psi)
\cdot \bm{p}_{K_f}
 F_{\psi \phi K_f,B}^{01} \ ,
\end{eqnarray}
where $c_{\rm dir}^{J^P}$ is a coupling constant
for the $J/\psi\phi(J^P)$ partial wave amplitude.

We basically use a common cutoff value (1~GeV unless otherwise stated)
in the form factors
for all the interaction vertices discussed
above.
One exception applies to Eqs.~(\ref{eq:dir_s0}) and (\ref{eq:dir_s1})
where we adjust $\Lambda^\prime$ of Eq.~(\ref{eq:ff1})
so that the $M_{J/\psi \phi}$ distribution from the direct decay amplitude
is similar to the phase-space shape.

In numerical calculations, 
for convenience, the above amplitudes 
are evaluated in the $J/\psi \phi$ center-of-mass frame.
An exception is the $Z_{cs}^{(\prime)}$ amplitudes that are 
evaluated in the total center-of-mass frame.
With the relevant kinematical factors multiplied to the amplitudes, 
the invariant amplitudes are obtained and plugged into the Dalitz plot
distribution formula;
see Appendix~B of Ref.~\cite{3pi} for details.

Parameter values obtained from and not from the fit are listed in
Tables~\ref{tab:param1} and \ref{tab:para2}, respectively.
In Table~\ref{tab:param1},
we also list each mechanism's fit fraction
defined by 
\begin{eqnarray}
\label{eq:ff}
{\rm FF.} = {\Gamma_{A_x} \over \Gamma_{\rm full}} \times 100\ (\%)\ ,
\end{eqnarray}
where $\Gamma_{\rm full}$ and $\Gamma_{A_x}$ are 
$B^+\to J/\psi\phi K^+$ decay rates calculated with the full model and
with an amplitude $A_x$ only, respectively. 
In Table~\ref{tab:param1}, 
$A^{\rm 1L}_{\psi'\phi(1^+)}$ seems to have a rather large fit fraction
of $\sim 36$\%.
This mechanism causes a threshold cusp at $M_{J/\psi\pi}\sim 4.7$~GeV, 
and its height (without interference) is about 80\% of the data.
The mechanism also has a long tail toward the lower $M_{J/\psi\pi}$
region, which makes its fit fraction rather large.
The amplitudes $A_{\rm dir}^{0^+}$ and $A_{\rm dir}^{1^+}$ also have large fit fractions of 
$\sim 67$\% and $\sim 45$\%, respectively.
This may be because $K^{(*)}_{J}$-excitation mechanisms
have been subsumed in this mechanism.
The LHCb analysis~\cite{lhcb_phi} found large fit fractions of 
the $K^{(*)}_{J}$-excitation mechanisms.

\section{Double triangle amplitudes and closeness to the leading
 singularity}
\label{app2}

The Coleman-Norton theorem~\cite{coleman} states that
a DT amplitude like Fig.~\ref{fig:diag}(a)
has the leading singularity
if the whole DT process is kinematically allowed at the
classical level:
the energy and momentum are always conserved;
in Fig.~\ref{fig:diag}(a),
all internal momenta are collinear
in the $D_{sJ}^{(*)}\bar{D}_{s}^{(*)}$ center-of-mass frame;
$D^{(*)}$ and $\bar{D}_{s}^{(*)}$ 
($\bar{K}$ and $K$) 
are moving to the same direction and the former is faster than the latter.

The DT amplitudes presented in Eqs.~(\ref{eq:DT1})-(\ref{eq:DT7})
do not exactly satisfy the above kinematical condition, and thus do not have the leading singularity.
Yet, their threshold cusps are significantly enhanced compared with
an ordinary one-loop threshold cusp.
This is because
the DT amplitudes are fairly close to satisfying the kinematical condition of the
leading singularity, and here we examine how close.

\begin{figure}[t]
\begin{center}
\includegraphics[width=.5\textwidth]{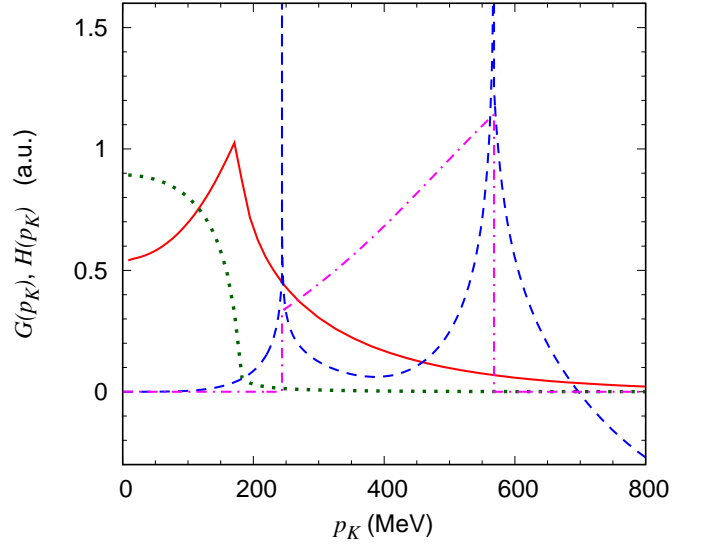}
\end{center}
 \caption{
$G$ and $H$ from the DT amplitude including the $p$-wave
 $D_{s1}\bar{D}_s$ pair, as defined in Eqs.~(\ref{eq:DT4a})-(\ref{eq:H});
$W=m_{D_{s1}}+m_{\bar{D}_s}+1.4~{\rm MeV}\sim 4505$~MeV.
The real and imaginary parts of 
$G$ [$H$] are shown by the 
blue dashed and magenta dash-dotted
[red solid and green dotted]
curves, respectively.
The relative magnitude between
$G$ and $H$ is arbitrary scaled to fit in the same figure.
 }
\label{fig:LS}
\end{figure}

Let us study
$A^{\rm DT}_{D_{s1}\bar{D}_s(0^+)}$ of Eq.~(\ref{eq:DT4}) 
that generates an $X(4500)$-like threshold cusp.
Apart from the coupling constants
and the dependence on the external $K^+$, we can express Eq.~(\ref{eq:DT4}) 
as
\begin{eqnarray}
 \label{eq:DT4a}
A^{\rm DT}_{D_{s1}\bar{D}_s(0^+)} &=&
\int d p_{K}\, G(p_{K}) H(p_{K})  \ ,
\end{eqnarray} 
with
\begin{eqnarray}
 \label{eq:G}
G(p_{K}) &=&  \int d\Omega_{p_{K}}
{p^2_{K}\bm{p}_{\bar{K}K}\!\cdot \bm{\epsilon}_\phi\,
\bm{p}_{\bar{K}\psi}\!\cdot\bm{\epsilon}_{\psi} f_{K\bar{K},\phi}^{1} 
 f_{\psi\bar{K}}^{1}
\over W-E_{K}-E_{\bar{K}}-E_{\psi}+i\epsilon}  , \\
 \label{eq:H}
H(p_{K}) &=&\int d^3p_{\bar{D}_s}
{
\bm{p}_{D^*\bar{D}_s}\cdot\bm{p}_{\bar{D}_s}
\over W-E_{K}-E_{D^*}-E_{\bar{D}_s}+i\epsilon} \nonumber\\
&&\times {f_{D^*\bar{D}_s}^{1}
 f_{D^*K,D_{s1}}^{0}
 f_{D_{s1}\bar{D}_s}^{1}
\over W-E_{D_{s1}}-E_{\bar{D}_s}+ {i\over 2} \Gamma_{D_{s1}}} ,
\end{eqnarray}
where $G(p_{K})$ and $H(p_{K})$
 have been implicitly projected onto $0^+$ of the
$J/\psi\phi$ pair.
Here, we suppose that $G(p_{K})$ and $H(p_{K})$ include only
$K=K^+$, $\bar{K}=K^-$, $D^*=D^{*0}$ in the two-loop, although their isospin
partners are also included in Eq.~(\ref{eq:DT4}).
Now we plot in Fig.~\ref{fig:LS} $G(p_{K})$ and $H(p_{K})$ for 
$W=m_{D_{s1}}+m_{\bar{D}_s}+1.4~{\rm MeV}\sim 4505$~MeV where the DT
amplitude is close to causing the leading singularity.
Both $G(p_{K})$ and $H(p_{K})$ show singular behaviors.
The real part of $H(p_{K})$ [red solid curve] shows a peak at 
$p_K\sim 170$~MeV (peak A). 
The peak A is due to a triangle singularity from the
$D_{s1}^+{D}^-_sD^{*0}$ triangle loop; see Fig.~\ref{fig:diag}(a).
Meanwhile, the angular integral of 
the $K\bar{K}J/\psi$ energy denominator in 
$G(p_{K})$ causes 
a logarithmic end-point singularity.
Thus the real part of $G(p_{K})$ [blue dashed curve] shows 
two peaks, and the one at $p_K\sim 240$~MeV (peak B) is
relevant to the leading singularity.
The other peak does not create a singular behavior in the amplitude of 
Eq.~(\ref{eq:DT4a}).
If the peaks A and B occurred at the same $p_K$,
the DT leading singularity (pinch singularity) would have occurred.
Yet, Fig.~\ref{fig:LS} indicates a substantial overlap between
the peaks A and B, and this is the cause of
the enhancement of the DT threshold cusps. 
The proximity of $A^{\rm DT}_{D_{s1}\bar{D}_s(0^+)}$
to the leading singularity condition is due to the fact that:
(i) each vertex is kinematically allowed to occur at on-shell;
(ii) the mass deficit in $D^{*}D_s^-\to \bar{K} J/\psi$ enables the relatively
light $\bar{K}$ to chase $K$ with a velocity faster than $K$.
In fact, if the exchanged 
$K^-$ mass were in the range of 
445 $\ltap m_{K^-} \ltap 455$~MeV,
$A^{\rm DT}_{D_{s1}\bar{D}_s(0^+)}$ would have hit the leading singularity.

%



\end{document}